%% file: paper.tex
\documentclass[10pt,conference]{IEEEtran}
\IEEEoverridecommandlockouts
\usepackage{cite}
\usepackage{amsmath,amssymb,amsfonts}
\usepackage{algorithmic}
\usepackage{graphicx}
\usepackage{textcomp}
\usepackage{xcolor}
\def\BibTeX{{\rm B\kern-.05em{\sc i\kern-.025em b}\kern-.08em
    T\kern-.1667em\lower.7ex\hbox{E}\kern-.125emX}}

\definecolor{bg}{rgb}{1.0,1.0,1.0}
\usepackage[frozencache,cachedir=.]{minted}
\usepackage{subcaption}

\usepackage{hyperref}
\hypersetup{
    colorlinks,
    linkcolor={red!50!black},
    citecolor={blue!50!black},
    urlcolor={blue!80!black}
}

\begin{document}

% Conducting Automated, 
%\title{Reusable, Test-driven Software Experiments with the LASSO Platform By Example: Assessing Code LLMs for Code Generation Reliability\\
\title{Test-driven Software Experimentation with LASSO: an LLM Prompt Benchmarking Example\\
%{\footnotesize \textsuperscript{*}Note: Sub-titles are not captured for https://ieeexplore.ieee.org  and
}

\author{\IEEEauthorblockN{Marcus Kessel}
\IEEEauthorblockA{
\textit{University of Mannheim}\\
Mannheim, Germany \\
0000-0003-3088-2166}
}

\maketitle

\begin{abstract}
Empirical software engineering faces a critical gap: the lack of standardized tools for rapid development and execution of Test-Driven Software Experiments (TDSEs) -- that is, experiments that involve the execution of software subjects and the observation and analysis of their ``de facto'' run-time behavior. In this paper we present a general-purpose analysis platform called LASSO that provides a minimal set of domain-specific languages and data structures to conduct TDSEs. By empowering users with an executable scripting language to design and execute TDSEs, LASSO enables efficient evaluation of run-time semantics and execution characteristics in addition to statically determined properties. We present an example TDSE that demonstrates the practical benefits of LASSO's scripting capabilities for assessing the reliability of LLMs for code generation by means of a self-contained, reusable and extensible study script. The LASSO platform and live pipeline examples are publicly available at: \url{https://softwareobservatorium.github.io/}.
%, and a demo video is available on YouTube: \url{https://youtu.be/tzY9oNTWXzw}.

% , empowering users to rapidly develop and execute TDSEs in software engineering

%Key features of LASSO include search and selection of exhibited behavior, as well as an integrated arena to obtain observational data for data-driven comparison using Jupyter notebooks. 
\end{abstract}


%% sections
\input{introduction} 
\input{scenario}
\input{extension}
\input{conclusion}

%\section*{Acknowledgment}

%This research is supported by the Ministry of Science, Research and Arts Baden-Württemberg, Research Seed Capital.

\bibliographystyle{IEEEtran}
\bibliography{IEEEabrv,literature}

\end{document}

%% file: introduction.tex
% FIXME mention LSLFlow -- or own paper on LSLFlow and TDSEHub?
% FIXME also point out new version of LSL - documentation online

\section{Introduction}
\label{sec:introduction}

Test-Driven Software Experiments (TDSEs) are controlled experiments that evaluate software artifacts, such as code modules, under various execution conditions \cite{kesselOS2023}. They reveal critical properties of run-time behavior that cannot be inferred solely through static analysis, as described by Rice's theorem \cite{riceTheorem}. TDSEs are widely used across software engineering: researchers use them to empirically validate tools and techniques involving code execution (e.g., code and test generation \cite{chen2021evaluatinglargelanguagemodels,10.1145/2685612}); practitioners apply them to evaluate tools for adoption (e.g., code recommendation \cite{kesselNextGen2023}); and educators employ them in programming courses to create interactive, feedback-rich exercises (e.g., \cite{10.1145/3639474.3640054}).

Despite their widespread use, designing, executing, and evaluating TDSEs remains a labor-intensive and often ad hoc process. Making software subjects testable, specifying reusable and meaningful tests, and capturing relevant observations in a controlled setting all present significant challenges \cite{kesselOS2023}. Consequently, current approaches -- typically based on general-purpose programming languages and custom scripts -- are difficult to scale, reuse, and reproduce.

To address these limitations, we developed \textbf{LASSO}, the \textit{Large-Scale Software Observatorium} \cite{madoc64107,kesselOS2023}. LASSO is a unified platform for creating automated, reproducible TDSEs at scale. Rather than relying on ad hoc scripting, LASSO provides domain-specific languages (DSLs) and tailored data structures that simplify the construction of executable study pipelines. Its scripting language, \textbf{LSL}, enables users to quickly and reliably define workflows that capture both dynamic (e.g., run-time semantics) and static (e.g., size-based metrics) properties.

LASSO has been used to support two major goals: (1) powering a suite of analysis services, including test generation \cite{kesselDivGen2022} and test-driven code recommendation \cite{kesselNextGen2023}; and (2) conducting large-scale TDSEs, such as evaluating behavior sampling techniques \cite{kesselBehaviorSampling2019,kesselNextGen2023} and replicating the HumanEval benchmark for code LLMs in MultiPL-E \cite{kesselOS2023}.

This paper showcases a new class of TDSE pipelines developed with LASSO, specifically focused on evaluating \textbf{prompting techniques} for code generation with large language models (LLMs). Prompting has emerged as a central technique for leveraging LLMs without costly fine-tuning \cite{10.1145/3695988}. It is used not only in code generation, but also in tasks such as test and oracle generation \cite{10765033,10.1145/3715107}, and is increasingly applied in research \cite{10795375}, practice as well as education.

% FIXME references
However, effective prompting\footnote{This includes ``hybrid'' approaches that use prompting in some ways as part of their workflows as well, e.g. HITS \cite{10.1145/3691620.3695501}} remains an open research challenge \cite{10.1145/3722108}: Which techniques work best for which tasks? How reusable are prompts across models? How many exemplars are needed for better performance? Can prompt templates be transferred between tasks? Our work addresses these questions by presenting executable prompting pipelines written in LSL that facilitate systematic TDSEs of prompting strategies, with functional correctness as the central evaluation criterion.

While prior work has introduced LASSO and some of its analysis services (e.g., curated executable corpora \cite{kesselCuratedDataSets2019}, test-driven code search \cite{kesselNextGen2023} or TDSEs in general \cite{kesselBehaviorSampling2019,kesselNextGen2023}), this paper focuses on novel TDSE pipelines that are specifically designed to support reproducible studies of code generation via prompting.

For this, the pipeline language for TDSE has been further evolved (i.e., extended) and simplified. To encourage broader adoption, the LASSO platform is publicly available at:

\begin{itemize}
    \item \url{https://softwareobservatorium.github.io/}
\end{itemize}

There, users can browse example LSL pipelines via \textit{TDSEHub} and experiment with real-world executable corpora as well as code LLMs (both locally and remotely using a demo instance). Instead of manually writing LSL scripts, we also provide frontends based on predefined script templates and scenarios to let users build and customize their own prompting scenarios they want to assess.

Section~\ref{sec:scenario} presents a concrete TDSE prompting scenario with an example LSL script, demonstrating key features of TDSE pipelines in LASSO. Section~\ref{sec:extensibility} discusses extensibility points for customizing studies. Section~\ref{sec:conclusion} concludes with reflections and future directions.

%% file: scenario.tex
\section{Example TDSE — Prompting Code LLMs}
\label{sec:scenario}

This section demonstrates the capabilities of the LASSO platform from the perspective of a user aiming to design and execute a TDSE using its scripting language, LSL. At its core, LASSO provides a scalable workflow engine powered by LSL, an imperative/declarative DSL based on Groovy. LSL draws inspiration from build management, continuous integration, and data mining languages, and enables modular, executable study pipelines (i.e., workflows) that define all essential analysis steps of TDSEs. For a detailed account of LASSO's architecture and semantics, and additional differential testing activities we refer to \cite{madoc64107,kesselOS2023,kesselGAI2024}.

\subsection{Study Design Overview} 

Suppose a user wishes to evaluate the performance of a specific prompting technique -- few-shot prompting (in-context learning) -- using a code LLM for the natural-language-to-code generation task. As a simplification, the objective is to sample $5$ Java solutions for a single coding task (Base64 encoding\footnote{Base64 encoding converts binary data into a string of printable ASCII characters, primarily used to transmit binary data (like images or audio) over channels that only support text}), described via a prompt in natural language. Prompting performance is assessed by checking whether the generated solutions pass representative input-output tests that reflect the desired functional behavior (i.e., satisfying functional correctness), following principles similar to existing LLM benchmarks like HumanEval \cite{kesselOS2023}. Oracle values are used to verify correctness in terms of expected outputs.

\subsection{Mapping the Design to LSL}

\begin{figure*}
  \begin{minipage}[t]{0.8\textwidth}
    \begin{minipage}{\textwidth}
      \input{script}
    \end{minipage}
      \vspace{-0.8cm}
    \caption{LSL Script}
    \label{fig:script}
    \vspace{0.3cm}
  \end{minipage}%  <-- Important: This prevents extra space
  \begin{minipage}[t]{0.2\textwidth}
    \centering
    \includegraphics[scale=0.22,trim=0cm 0cm 0cm 0cm,clip]{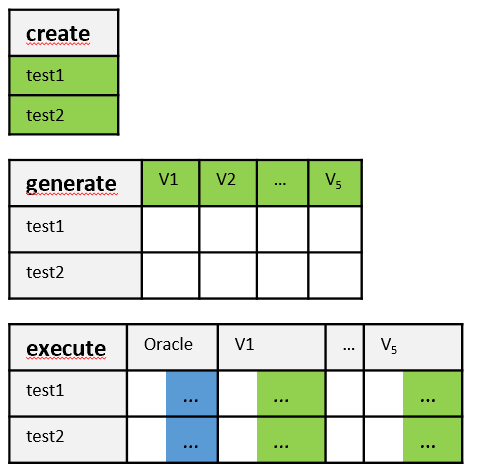}
    \caption{Evolution of SRMs over Actions}
    \label{fig:srm}
    \vspace{10ex}
    \includegraphics[scale=0.25,trim=0cm 0cm 0cm 0cm,clip]{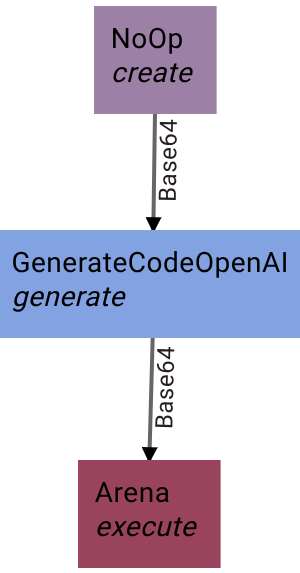}
    \caption{Action Execution Graph}
    \label{fig:dag}
  \end{minipage}
    \vspace{-0.3cm}
  \caption{TDSE Pipeline Script for Prompt Assessment -- Generating 5 Code Solutions Using GPT-4o-mini}
  \label{fig:lsl_script_description}
  \vspace{-0.5cm}
\end{figure*}

Figure~\ref{fig:script} shows how the study design is translated into an executable LSL script that showcases LASSO's core features.

The study involves three core steps --

\begin{enumerate}
    \item Preparing the coding problem (Base64 encoding) with its interface and tests.
    \item Prompting a code LLM (OpenAI's GPT-4o-mini) to generate five implementations.
    \item Executing the implementations to assess their behavior against the two test cases.
\end{enumerate}

These are reflected in the LSL actions \texttt{create}, \texttt{generate}, and \texttt{execute} respectively (see lines 15, 32, and 52 in Figure~\ref{fig:script}). The generated implementations are evaluated using LASSO's scalable test engine (``arena''), which executes the test suite and captures observations for further analysis (e.g., behavorial clustering, correctness checks).

\subsubsection{Reusable Study Pipelines with Actions}

LSL pipelines are structured as sequences of reusable \textit{actions}, which represent composable units of analysis steps. These actions define data dependencies and execution order, and can be nested to create rich, multi-step workflows.

While LASSO allows flexible granularity in action design, most study pipelines consist of two major types --

\begin{itemize}
    \item \textbf{Generation actions}, which produce or retrieve code modules (e.g., via LLMs, code corpus search or other data sources).
    \item \textbf{Execution actions}, which observe, measure and analyze code behavior (e.g., via test drivers and measurement calipers).
\end{itemize}

The generated code module implementations are added to LASSO's executable corpus -- a curated collection of real-world code artifacts, enabling both textual and test-driven code retrieval (i.e., behavior sampling) for other use cases such as code recommendation \cite{kesselNextGen2023}.

\subsubsection{Specifying Functional Behavior}

Each coding task in LASSO is defined through a \textit{functional abstraction}, which captures the expected behavior using (1) an interface specification (defined in LQL, LASSO's query language), and (2) a test suite of input-output examples.

These are bundled into abstraction containers (see line 17-29) that flow between actions (e.g., from \texttt{generate} to \texttt{execute}). In our example, the Base64 problem is specified to include padding (e.g., using ``='' characters as per RFC 4648\footnote{See \url{https://www.rfc-editor.org/rfc/rfc4648}}; see line 46), with inputs and expected outputs defined in a concise tabular format using LSL's sequence sheet notation syntax (SSN) (lines 23-28).

\paragraph{Interface Signatures in LQL}

LQL enables expressive yet succinct definitions of input-output signatures (e.g., class interface and method signatures in Java, or modules and functions in Python). Tests reference these signatures rather than concrete implementations, allowing LASSO to automatically synthesize adapters when needed. This ensures consistent testability across heterogeneous code solutions.

\paragraph{Data Structures for Behavioral Observation}

LASSO uses three hierarchically-nested data structures\footnote{Technical documentation is available here: \url{https://softwareobservatorium.github.io/web/docs/category/data-structures-and-languages}} to capture run-time behavior --

\begin{itemize}
    \item \textbf{Sequence Sheet Notation}: Tabular test definitions in LSL or as spreadsheets (JSONL format).
    \item \textbf{Stimulus-Response Matrices (SRMs)}: Aggregate test results and arbitrary observations across code implementations.
    \item \textbf{Stimulus-Response Hypercubes (SRHs)}: Multi-dimensional collections of SRMs capturing broader experimental variations.
\end{itemize}

These abstractions allow scalable behavioral analysis across many TDSE scenarios (see \cite{kesselOS2023}). Figure~\ref{fig:srm} shows the evolution of SRMs over multiple actions. The process begins with the \texttt{create} action, where an initial stimulus matrix is generated by adding rows from two tests. Next, in the \texttt{generate} action, up to 5 code candidate implementations generated from the code model are added as columns. Finally, the \texttt{execute} action stores the output observations in the corresponding cell of the SRM, which can then be stored and analyzed in the SRH.

\subsection{Script Execution and Environments}

Figure~\ref{fig:dag} shows the script's directed, acyclic graph (DAG), derived automatically by LASSO's engine. It visualizes dependencies among actions and the flow of abstraction containers.

Prompting of code models is natively supported by the corresponding type within the \texttt{generate} action. As shown in Figure~\ref{fig:script}, (Groovy's) template strings can be utilized to create dynamic prompts tailored to the pipeline's underlying data model (i.e., the placeholder \texttt{\$\{stimulusMatrix.lql\}} in the example script, which holds the LQL interface specification of the current abstraction container). Furthermore, both the action's configuration options and the prompt itself are customizable, allowing users to tailor the \texttt{generate} action to their specific needs.

LASSO's arena executes the test suite for each code candidate implementation in a controlled Docker environment (configured in line 4) that can be individually configured. The execution uses Java JDK 17 (Eclipse Temurin), and results are recorded in SRMs. LASSO supports both vertical (multi-threaded) and horizontal (multi-node) scaling, but can also run in standalone (embedded) mode on a single machine.

\subsection{Analyzing Observed Behavior}

LASSO supports two complementary analysis approaches:

1. \textbf{Script-driven analysis}, embedded directly into the LSL pipeline via DSL commands.
2. \textbf{Data-driven analysis}, performed externally using analytics tools (e.g., Python Pandas with Jupyter/Google Colab) on exported SRMs/SRHs (as CSV or Parquet).

\begin{figure*}[h]
  \begin{subfigure}[t]{\textwidth}
    \centering
    \includegraphics[scale=0.12]{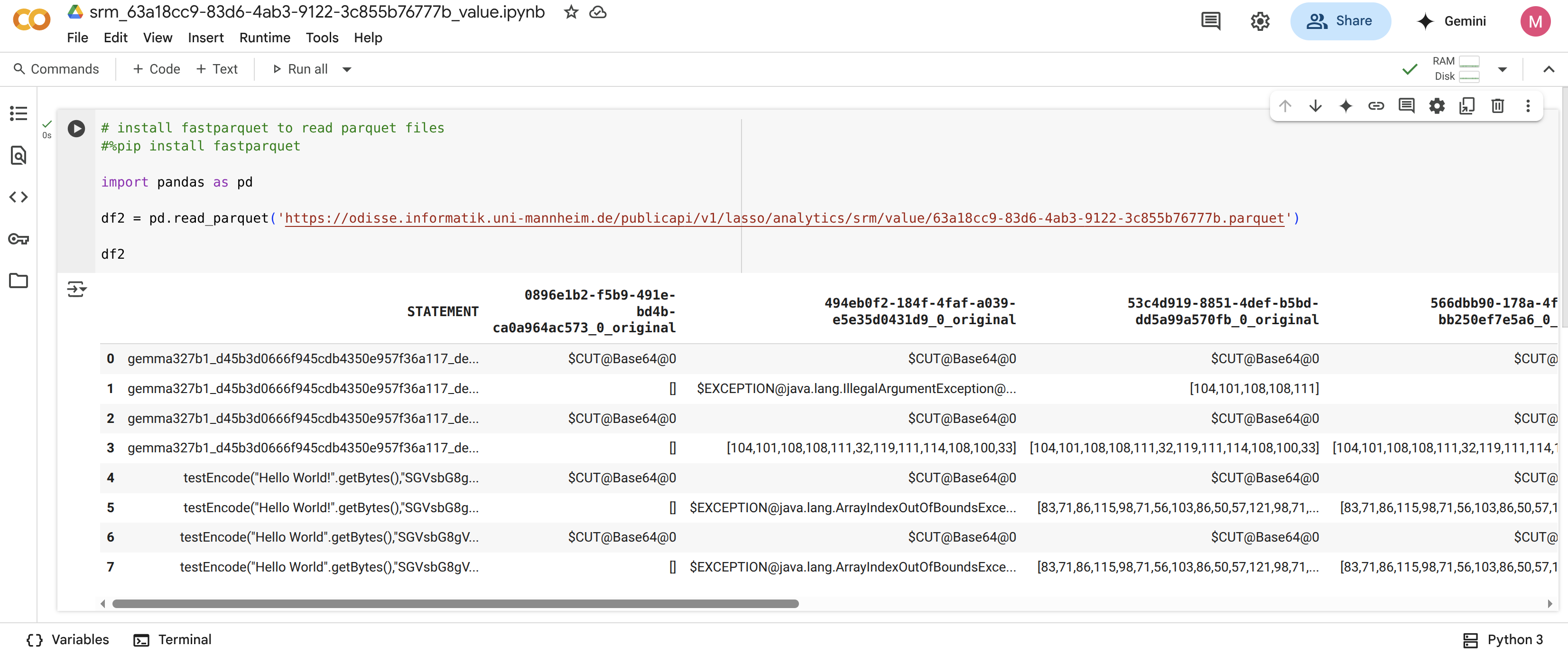}
    \caption{Interactive Data-driven Analysis of SRMs in Notebooks (here using Google Colab)}
    \label{fig:analytics}
    \vspace{0.5cm}

    \includegraphics[scale=0.14,trim=0cm 0cm 0cm 0cm,clip]{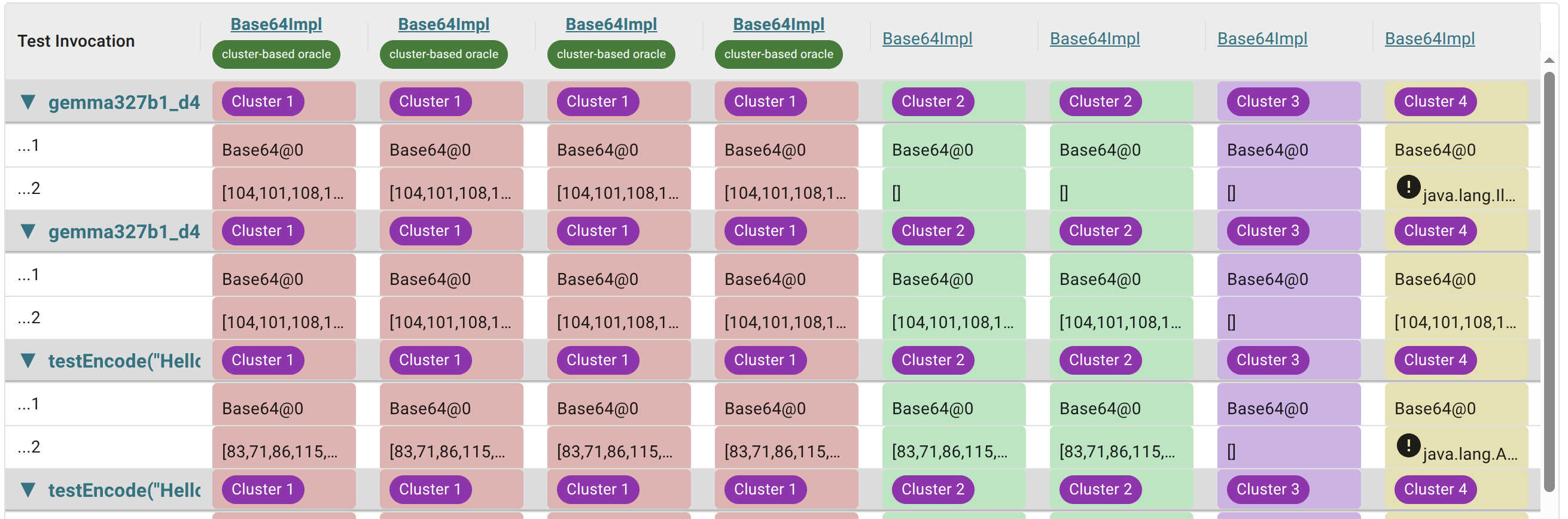}
    \caption{Behavorial Clustering}
    \label{fig:clustering}
  \end{subfigure}
  \caption{SRM-based Behavioral Clustering}
  \label{fig:srm_analysis}
  \vspace{-0.5cm}
\end{figure*}

The latter is often preferred, as it decouples execution from evaluation and allows greater flexibility. Figure~\ref{fig:analytics} illustrates this post-execution analysis flow in Google Colab (similar to Jupyter Notebooks).

In SRMs, tests form rows and implementations form columns. Oracle values appear as virtual reference columns (see first column in Figure~\ref{fig:srm}). Functional equivalence is identified via pairwise comparisons of the columns (i.e., code candidate implementations). LASSO also offers frontend support for behavioral clustering (see Figure~\ref{fig:clustering}), enabling users to detect discrepancies via differential testing \cite{kesselGAI2024}.

TDSEs can be iteratively refined by re-running scripts with modified prompts, added tests, or expanded coding problems, gradually building a comprehensive picture of LLM behavior and reliability. The SRMs resulting from different prompts, for instance, can then be compared applying differential testing techniques to their SRH representation, in order to assess their performance (i.e., increase/decrease in functional correctness on average).

%% file: script.tex
\begin{minted}[escapeinside=!!,breaklines,fontsize=\scriptsize,bgcolor=bg,linenos]{groovy}
dataSource 'lasso_quickstart'
study(name: 'ChatGPT') {

    profile('java17Profile') { // target execution profile
        scope('class') { type = 'class' }
        environment('java17') {
            image = 'maven:3.9-eclipse-temurin-17'
        }
    }

    /** 
    * Initialize stimulus matrix with tests
    * (note: tests are typically provided through frontends or are generated/harvested)
    */
    action(name: 'create') {
        execute {
            stimulusMatrix('base64', """
Base64 {
    encode(byte[])->byte[]
    decode(java.lang.String)->byte[]
}
            """, [/*impls: empty list*/], [
            test(name: 'testEncode(p1=byte[], p2=byte[])', p1:'"Hello World!".getBytes()', p2:'"SGVsbG8gV29ybGQh".getBytes()') {
                row '', 'create', 'Base64'
                row '?p2', 'encode', 'A1', '?p1'
            },
            test(name: 'testEncode(p1=byte[], p2=byte[])', p1:'"Hello World".getBytes()', p2:'"SGVsbG8gV29ybGQ=".getBytes()')
            ])
        }
    }

    action(name: 'generate', type: 'GenerateCodeOpenAI') {
        dependsOn 'create'
        include 'base64'
        profile('java17Profile')

        // action configuration block
        apiKey = "demo"
        model = "gpt-4o-mini"
        samples = 5

        // custom DSL command: for each stimulus matrix, create one prompt to obtain impls
        prompt { stimulusMatrix ->
            // can by for any prompts: FA, impls, models etc.
            def prompt = [:] // create prompt model
            prompt.promptContent = """implement a java class with the following interface specification, but do not inherit a java interface: ```${stimulusMatrix.lql}```. The implementation must support Base64 padding. Only output the java class and nothing else."""
            prompt.id = "lql_prompt"
            return [prompt] // list of prompts is expected
        }
    }

    action(name: 'execute', type: 'Arena') {
        dependsOn 'generate'
        include 'base64'
        profile('java17Profile')
    }
}  
\end{minted}

%% file: extension.tex
\vspace{-0.2cm}
\section{Extensibility}
\label{sec:extensibility}

The self-contained nature of LSL scripts, as demonstrated in the previous section, provides several advantages beyond reproducibility. Most notably, by encoding study designs and assumptions explicitly as executable code, LSL scripts are highly reusable, modifiable, and easy to re-execute. Key parameters, such as the code LLMs under study, the number of code solutions to sample, the prompts and model-specific settings (e.g., temperature), can be defined as global variables in the script header, enabling users to adapt study designs with minimal effort.

LSL scripts also support the flexible definition of new coding problems. Users can simply add new abstraction blocks to define problems ad hoc or reuse existing datasets and benchmarks like HumanEval, as demonstrated in the replicated study from \cite{kesselOS2023}, to load predefined coding problem specifications.

Furthermore, the modular structure of LSL pipelines facilitates seamless integration of additional actions. Users can extend experiments by incorporating new steps, methods, or tools, whether to refine existing analyses or to explore new research questions.

\begin{listing}
    \input{test_script}
    \caption{Extending TDSE Pipelines: Integrating Ollama-based LLMs and Test Generation via LLMs and EvoSuite}
    \label{fig:lsl_script_extensions}
\end{listing}

As shown in Listing~\ref{fig:lsl_script_extensions}, researchers can expand a TDSE by integrating additional LLMs, such as locally hosted models via Ollama\footnote{\url{https://ollama.com/}}, through the inclusion of dedicated pipeline actions. Similarly, users seeking to improve confidence in functional correctness may introduce new test generation actions, leveraging LLM-based methods or established tools like EvoSuite \cite{10.1145/2685612}.

Finally, LASSO's extensibility includes support for custom user-defined actions. In addition to the platform's default set of documented actions (available via the LASSO project website), users can implement their own analysis steps using the platform's well-defined Actions API. These custom actions can be packaged and deployed via Docker containers, enabling seamless integration into the workflow engine.

In summary, LSL scripts offer a robust foundation for extensible TDSEs, fostering reproducibility, reusability, and modular experimentation in code-related empirical studies.

%% file: test_script.tex
\begin{minted}[escapeinside=!!,breaklines,fontsize=\scriptsize,bgcolor=bg]{groovy}
/* use LLMs supported by Ollama: here deepseek-r1:latest */
action(name: 'generateCodeDeepSeek', type: 'GenerateCodeOllama') {
    ...

    // action configuration block
    servers = ["http://localhost:11434"]
    model = "deepseek-r1:latest"
    samples = 5

    // custom DSL command offered by the action (for each stimulus matrix, create one prompt to obtain impls)
    prompt { stimulusMatrix ->
        // can by for any prompts: FA, impls, models etc.
        def prompt = [:] // create prompt model
        prompt.promptContent = """write a java class based named 'Base64Impl' on the following interface specification, but do not inherit a java interface: ```${stimulusMatrix.lql}```. Only output the java class and nothing else."""
        prompt.id = "lql_prompt"
        return [prompt] // list of prompts is expected
    }
}

/* generate additional tests using LLMs */
action(name: 'generateTestsLlama', type: 'GenerateTestsOllama') {
    ...

    // action configuration block
    servers = ollamaServers
    model = "gemma3:27b"
    samples = 1

    prompt { stimulusMatrix ->
        def prompt = [:] // create prompt model
        prompt.promptContent = """write a junit test class to test the functionality of the following interface specification: ```${stimulusMatrix.lql}```. Assume that the specification is encapsulated in a class that uses the same naming as in the interface specification. Only output the JUnit test class and nothing else."""
        prompt.id = "lql_prompt"
        return [prompt] // list of prompts is expected
    }
}

/* generate tests using EvoSuite */
action(name: 'evoSuite', type: 'EvoSuite') {
    // action configuration block
    searchBudget = 120 // we need this as upper bound for timeouts
    stoppingCondition = "MaxTime"

    ...
}
\end{minted}

%% file: conclusion.tex
\section{Conclusion}
\label{sec:conclusion}

This paper has presented LASSO, a scalable platform for automating test-driven software experiments (TDSEs), and LSL, its domain-specific scripting language for specifying reusable and modular study pipelines. Using a concrete scenario -- evaluating prompting techniques for code generation with LLMs -- we demonstrated how users can translate study designs into executable LSL scripts that define, generate, and analyze experiments systematically. By combining code generation, differential testing, and data-driven behavioral analysis, LASSO enables researchers to explore the functional correctness of generated code solutions at scale.

We also highlighted how LSL pipelines support extensibility through modular action blocks, making it easy to integrate additional code models, test generators, and even user-defined tools. These capabilities foster reproducibility and promote iterative refinement in TDSEs.

To support community adoption, we have created \textsc{TDSEHub}, a curated and growing collection of LSL-based TDSE pipelines. TDSEHub serves as a central resource for researchers and practitioners to explore example studies, reuse existing designs, and understand LASSO's capabilities. In the mid-term, we plan to expand TDSEHub into a collaborative space where users can contribute their own pipelines and share experimental results.

In addition to its support for Java, LASSO now provides initial support for the Python programming language, including an arena test driver for Python code and preliminary analyzers. These enhancements are part of our ongoing work to broaden language support and lower the barrier for adopting TDSEs across different domains and toolchains.

LASSO is implemented in Java using Spring Boot and can be deployed as a standalone application or across distributed environments using Docker. A web-based dashboard and API allow users to submit, manage, and execute LSL scripts efficiently.

By lowering the barrier to authoring, executing, and analyzing test-driven software experiments, LASSO and LSL pave the way for a more reproducible and collaborative future in empirical software engineering.

% FIXME documentation of DSL, pipelines etc.

The LASSO platform, documentation, example TDSEs and frontends to its services are publicly available at:

\begin{itemize}
    \item \url{https://softwareobservatorium.github.io/}
\end{itemize}